\documentclass[
  twocolumn,
  prb,
  amsmath,
  amssymb,
  superscriptaddress,
  floatfix
]{revtex4}

\usepackage{bm}
\usepackage{graphicx}

\begin{document}

\title{Symmetries and weak (anti)localization of Dirac fermions in HgTe quantum wells
}

\author{P.\ M.\ Ostrovsky}

\affiliation{Max-Planck-Institut f\"ur Festk\"orperforschung, Heisenbergstr. 1,
70569, Stuttgart, Germany}

\affiliation{
 Institut f\"ur Nanotechnologie, Karlsruhe Institute of Technology,
 76021 Karlsruhe, Germany
}
\affiliation{
 L.~D.~Landau Institute for Theoretical Physics RAS,
 119334 Moscow, Russia
}

\author{I.\ V.\ Gornyi}

\affiliation{
 Institut f\"ur Nanotechnologie, Karlsruhe Institute of Technology,
 76021 Karlsruhe, Germany
}
\affiliation{
 A.F.~Ioffe Physico-Technical Institute,
 194021 St.~Petersburg, Russia.
}
\affiliation{
 DFG Center for Functional Nanostructures,
 Karlsruhe Institute of Technology, 76128 Karlsruhe, Germany
}

\author{A.\ D.\ Mirlin}

\affiliation{
 Institut f\"ur Nanotechnologie, Karlsruhe Institute of Technology,
 76021 Karlsruhe, Germany
}
\affiliation{
 Inst. f\"ur Theorie der kondensierten Materie,
 Karlsruhe Institute of Technology, 76128 Karlsruhe, Germany
}
\affiliation{
 DFG Center for Functional Nanostructures, Karlsruhe Institute of Technology,
 76128 Karlsruhe, Germany
}

\affiliation{
 Petersburg Nuclear Physics Institute,
 188300 St.~Petersburg, Russia.
}

\begin{abstract}

We perform a symmetry analysis of a 2D electron system in HgTe/HgCdTe quantum
wells in the situation when the chemical potential is outside of the gap, so
that the bulk of the quantum well is conducting. In order to investigate quantum
transport properties of the system, we explore
symmetries of the low-energy Hamiltonian which is expressed in
terms of two flavors of Dirac fermions, and physically important
symmetry-breaking mechanisms. This allows us to predict emerging patterns of
symmetry breaking that control the weak localization and antilocalization
showing up in transverse-field magnetoresistance. 

\end{abstract}

\maketitle

\section{Introduction}
\label{s1}

Solid state systems with massless Dirac charge carriers have recently 
attracted an outstanding degree of attention of theoretical and
experimental groups worldwide. 
This is mainly due to the impressive experimental progress in two 
physically related directions of research: graphene \cite{CastroNeto09,nobel}
and topological insulators \cite{Hasan10RMP,Qi11}.

The notion of a topological insulator (TI) refers to a bulk insulator with 
gapless surface states occurring due to topological reasons. The
simplest 
example of a topological insulator is the 2D electron gas in a strong magnetic 
field. At the quantum Hall plateau, the gap between Landau levels in the bulk is 
penetrated by a fixed integer number of chiral edge states providing the quantized 
value of the Hall conductance.
The integer quantum Hall edge is thus a topologically protected
1D conductor realizing the group $Z$.  

A novel class of TIs \cite{kane,BernevigHughesZhang,Koenig07,Fu07,hasan,Hasan10RMP,Qi11} requires
strong spin-orbit interaction in the absence of magnetic field (i.e. it is realized
in systems with preserved time-reversal invariance). 
This type of TIs was discovered in
HgTe/HgCdTe structures by the W{\"u}rzburg group \cite{Koenig07}.  
Strong spin-orbit interaction in HgTe leads to the inverted band gap in this semiconductor.
As a result, the electron and hole bands are crossing near the boundary of the sample giving
rise to the two counter-propagating helical edge modes. The time-reversal symmetry of the
system leads to the topological protection of these edge modes.

Voltage applied to such a sample results in the appearance of the perpendicular spin current. 
This phenomenon 
is known as the quantum spin-Hall effect (QSHE). The robustness of the effect with respect 
to disorder makes it an extremely promising tool for applications. As a simplest example, 
the conversion between the usual charge current and spin current occurring in QSHE can be 
used for generation and detection of spin currents.
The existence of a non-localized conducting channel at the edge of an (appropriately
manufactured) two-dimensional HgTe/HgCdTe
quantum well was experimentally demonstrated in Refs. \onlinecite{Koenig07,Roth09}.
These experiments, showing that HgTe/HgCdTe quantum wells provide a realization of a novel remarkable
class of materials---$Z_2$ topological insulators---opened a new exciting
research direction.

A 3D realization of a $Z_2$ TI was discovered in the experiment by Princeton group \cite{hasan}
where crystals of Bi$_{1-x}$Sb$_{x}$ were studied (later other Bismuth compounds such as BiTe and BiSe
were also shown to be 3D topological insulators). 
The 3D topological insulators exhibit very strong spin-orbit interaction also leading 
to the band gap inversion. This results in the appearance of gapless states on the surface of the sample
forming a 2D topologically protected metal. 
The dynamics of the surface states is governed by the same massless Dirac Hamiltonian that 
has previously appeared in graphene. The main difference between graphene and the surface 
of a 3D topological insulator is the lack of the spin and valley degeneracy in the latter case.
A combined effect of topology, disorder and interaction on a surface of 3D
topological insulators
was demonstrated to establish a novel critical 2D state with the low-temperature 
conductivity of order unity.~\cite{ostrovsky10} 

A similar critical state is expected~\cite{ostrovsky10} to
separate the normal and topological insulator states in a 2D $Z_2$ topological insulators.
In particular, such critical state should occur exactly at the QSHE transition in HgTe/HgCdTe
quantum wells of critical thickness, when the 2D bulk gap is tuned to zero.
The properties of this topologically protected metal reflect the
quasirelativistic Dirac nature of carriers, similarly to the interference 
phenomena in graphene.\cite{McCann,Nestoklon}

A two-dimensional metallic state in HgTe/HgCdTe quantum wells can also be realized 
in the presence of the bulk gap (either on the TI on the normal insulator side)
by shifting the chemical potential (with a help of the gate) away from the gap.
Such two-dimensional TI away from the TI regime represents a 2D spin-orbit
metal, whose properties (in particular, interference corrections to the
conductivity and low-field magnetoresistance) may still reflect the
Dirac-fermion nature of carriers.
In recent transport experiments  the magnetoresistance of bulk-conducting
HgTe/HgCdTe 
quantum wells was studied, both for inverted (thick
wells)\cite{kvon,minkov,bruene-unpub} 
and normal (thin wells)\cite{kvon,bruene-unpub} band structures.

Although some aspects of the interference corrections in TI systems away from the TI regime
have been already addressed theoretically \cite{Tkachov11}
there is a clear need in a systematic theory describing the whole variety of
experimentally accessible regimes. The purpose of this paper is to develop
such a theory. Analyzing the underlying Dirac-type Hamiltonian and physically
relevant symmetry-breaking terms, we identify parameter regimes with different
symmetries. The corresponding symmetry-breaking patterns determine the form of
quantum interference magnetoresistance, i.e. the weak localization (WL) vs weak
antilocalization (WAL) and corresponding prefactors. 

The paper is organized as follows.
In Sec.~\ref{s2} we analyze symmetry properties of the Bernewig-Hughes-Zhang
Hamiltonian and various physically relevant symmetry-breaking perturbations.
This allows us to establish emerging patterns of symmetry breaking.
In Sec.~\ref{s3} we use these results to calculate the interference corrections
(weak localization and weak antilocalization magnetoresistance) in various parts
of the parameter space. Section \ref{s4} contains a summary of our results and
a brief comparison to available experimental data.

\section{Effective Hamiltonian and symmetries}
\label{s2}

The standard approach to the description of electronic states in semiconductors
is based on the $\mathbf{k} \cdot \mathbf{p}$ method. This method assumes that
exact wave functions in the periodic potential of the crystal are known at some
given point in the Brillouin zone. Normally, the states with zero quasimomentum
($\Gamma$ point) are used. Then the effective Hamiltonian is constructed in the
basis of these states by expansion in small momentum $\mathbf{k}$. The $\mathbf{k}
\cdot \mathbf{p}$ perturbation theory thus provides the model Hamiltonian for
the states close to the $\Gamma$ point in terms of few matrix elements
calculated with respect to exact eigenstates at $\mathbf{p} = 0$. The
symmetries of the underlying system impose a number of constraints on the
structure of the effective theory. In fact, a general form of the effective
2D Hamiltonian can be derived from the symmetry grounds, see Ref. \onlinecite{Rothe10},
without invoking the microscopic (tight-binding or Kane-model) Hamiltonian.

The only intrinsic symmetry of the 3D Hamiltonian is time-reversal (TR): $H =
s_y H^* s_y$. In the clean case, the point symmetry group is $T_d$
(tetrahedral). Spatial inversion interchanges Hg and Te sublattices of the
crystal. Neglecting the difference between Hg and Te, the point group becomes
$O_h$ (cubic). The approximate inversion symmetry leads to the
block-diagonal structure of the BHZ Hamiltonian.

\subsection{BHZ Hamiltonian}
\label{s2.1}

Effective Hamiltonian for a narrow symmetric HgTe quantum well (QW) was derived in
Ref.\ \onlinecite{BernevigHughesZhang} in the framework of the
$\mathbf{k}\cdot\mathbf{p}$ method. The Bernevig-Hughes-Zhang (BHZ)
Hamiltonian has a $4 \times 4$ matrix
structure in the spin (sign of the $z$-projection of the total momentum $\textbf{J}$,
where the $z$-axis is perpendicular to the QW plane)
and E1 -- H1 space,
\begin{gather}
 H_\text{BHZ}
  = \begin{pmatrix}
      h(\mathbf{k}) & 0 \\
      0 & h^*(-\mathbf{k})
    \end{pmatrix}, \\
 h(\mathbf{k})
  = \begin{pmatrix}
      \epsilon(\mathbf{k}) + m(\mathbf{k}) & A (k_x + i k_y) \\
      A (k_x - i k_y) & \epsilon(\mathbf{k}) - m(\mathbf{k})
    \end{pmatrix}.
    \label{1}
\end{gather}
Here we have used the form given in Refs. \onlinecite{Qi11,Liu08,Rothe10} with the following
arrangement of components in the spinor: $E1+,H1+,E1-,H1-$.
Introducing the Pauli matrices $\sigma_{0,x,y,z}$ for the E1---H1 space
and $s_{0,x,y,z}$ for the up---down (spin) space (with $\sigma_0$ and $s_0$ unity matrices),
the effective Hamiltonian can be written as
\begin{equation}
H_\text{BHZ} 
  = \epsilon(\mathbf{k})\sigma_0 s_0 + m(\mathbf{k}) \sigma_z s_0 + A
 k_x \sigma_x s_z - A k_y \sigma_y s_0.
\label{2}
\end{equation}

The functions $\epsilon(\mathbf{k})$ and $m(\mathbf{k})$ are effective energy
and mass. Within the $\mathbf{k}\cdot\mathbf{p}$ expansion in the vicinity of
the $\Gamma$ point, they are
\begin{equation}
 \epsilon(\mathbf{k})
  = C + D \mathbf{k}^2,
 \qquad
 m(\mathbf{k})
  = M + B \mathbf{k}^2.
  \label{3}
\end{equation}
The two phases of normal and topological insulator correspond to $M>0$ and
$M<0$, respectively. The sign of $M$ changes at the critical thickness $d_c$ of the
QW of about $6.2$ nm. \cite{Koenig07} Parameters $A$, $B$, and $D$ are positive with
$B > D$. The parameter $C$ includes chemical potential and can be varied by
changing the electron concentration with the gate voltage.

The Hamiltonian $H_\text{BHZ}$ breaks up into two blocks acting independently in
the spin-up and spin-down subspaces with the spectrum
\begin{equation}
 E_\pm(\mathbf{k})
  = \epsilon(\mathbf{k}) \pm \sqrt{A^2 k^2 + m^2(\mathbf{k})}.
  \label{4}
\end{equation}

The standard way \cite{Tkachov11} to introduce disorder in the model is to add the fully
diagonal term
\begin{equation}
 H_\text{dis}
  = V(\mathbf{r}) \sigma_0 s_0
  \label{5}
\end{equation}
with a random potential $V(\mathbf{r})$ to the effective Hamiltonian $H_\text{BHZ}$. This
corresponds to a relatively smooth (on the scale of the quantum well thickness) disorder.
In particular, this model correctly describes charged impurities
located at a certain (sufficiently large) distance $R$ from the QW, e.g., in the doping layer.
In this situation, within the QW the impurity potential is almost constant in $z$ direction
(across the QW), so that it does not break the $z\to -z$ symmetry of the QW. Alternatively, one
can consider, e.g., placing short-range impurities exactly in the middle of the QW: such impurities
would also preserve the $z \to -z$ symmetry and hence would not give rise to the
mixing of the up and down blocks in $H_\text{BHZ}$.

In general, the Hamiltonian $H_\text{BHZ} + H_\text{dis}$ possesses two
symmetries. Apart from the physical time-reversal symmetry $H(\mathbf{k}) = s_y
H^*(-\mathbf{k}) s_y$, that relates the two spin blocks (Kramers doublets), the Hamiltonian
commutes with $s_z$. This allows one to consider the two blocks separately.
An extra symmetry can arise at some specific values of energy. In particular,
$h(\mathbf{k})$ acquires its own time-reversal symmetry when the mass $m(k_F)$
is zero. This happens in the inverted regime ($M < 0$) when $M + B k_F^2 = 0$.

It is also possible to achieve approximate orthogonal symmetry close to the
bottom of the band, when $|M| \gg \{A k_F, B k_F^2\}$, and at very high energies
$B k_F^2 \gg \{|M|, A k_F\}$. In the absence of block mixing, this would lead to
WL. 
On the other hand, the
regions of WL behavior at stronger magnetic fields 
(crossing over to WAL in weaker fields) were found in an experiment on similar
structures \cite{minkov,kvon}, indicating that one can indeed achieve the
regime of (approximate) orthogonal symmetry in this class of devices.

Thus we have identified three possible relevant symmetries of the model:
physical time-reversal, spin, and block-wise time-reversal 
(either ``symplectic'' or ``orthogonal''). The latter symmetry
can occur only in the presence of the spin symmetry, when the Hamiltonian
breaks into two blocks. When all three symmetries are present, the two copies of
symplectic or orthogonal class are realized leading to double WAL 
(i.e., with magnitude twice larger than the usual WAL correction)
or double WL correction (with the same prefactor as in the conventional 
spinful orthogonal class).
If the block-wise time-reversal symmetry is broken while spin symmetry is
preserved, we obtain two copies of the unitary class with no interference
corrections (in the lowest one-loop order). If the spin symmetry is also broken,
the system realizes a single copy of symplectic class yielding the usual (not
double) WAL correction. Finally, if the physical time-reversal
symmetry is broken (e.g., by magnetic impurities), the single copy of unitary
class is realized with no interference corrections (in the lowest order).

\subsection{Symmetry breaking mechanisms in 2D}
\label{s2.2}

Let us list possible mechanisms of symmetry breaking in a 2D quantum well of
HgTe.

\textit{Block-wise time-reversal symmetry} is not exact from the very
beginning. It is exact for massless Dirac fermions but is violated
by the presence of $m(k)\neq 0$. In $H_{\text{BHZ}}$, this symmetry
occurs only at one particular electron concentration. Since
disorder mixes the states in the energy window of order $1/\tau$ near Fermi
surface, it will inevitably break this symmetry. Another, and more relevant,
possibility is the detuning of the average electron concentration from the point
$M + B k_F^2 = 0$.

\textit{Spin symmetry} is broken by one of the following mechanisms:

(i) The block-diagonal structure of
$H_\text{BHZ}$ is not exact. Off-diagonal elements in the effective
Hamiltonian arise due to bulk inversion asymmetry (BIA) of the zinc-blende lattice of
HgTe and have the form\cite{Liu08}
\begin{multline}
 H_\text{BIA}
  = \begin{pmatrix}
      0 & 0 & 2\delta_e k_+ & -\Delta_0 \\
      0 & 0 & \Delta_0 & 2\delta_h k_- \\
      2\delta_e k_- & \Delta_0 & 0 & 0 \\
      -\Delta_0 & 2\delta_h k_+ & 0 & 0
    \end{pmatrix}
   = \Delta_0 \sigma_y s_y\\
   + \delta_e(\sigma_0+\sigma_z)(k_x s_x - k_y s_y)
  + \delta_h (\sigma_0-\sigma_z)(k_x s_x+k_y s_y).
  \label{6}
\end{multline}
Here the first term with $\Delta_0$ comes from the $k_z^2$ matrix elements
connecting electrons from the $\Gamma_6$ band and heavy holes from the $\Gamma_8$
band that have opposite spins (such coupling is absent in a spherically symmetric
Kane model in a bulk system).\cite{Winkler}
This term leads to the splitting of the spectrum into four branches:
\begin{equation}
 E_\pm(\mathbf{k})
  = \epsilon(\mathbf{k}) \pm \sqrt{(A|k|\pm \Delta_0)^2 + m^2(\mathbf{k})}.
  \label{EDelta}
\end{equation}
The linear-in-$k$ terms with $\delta_e$ and $\delta_h$ arise
after the projection  of the bulk cubic Dresselhaus terms for electrons
and holes, respectively, onto the QW.
Typically, the spin-orbit interaction for holes is stronger
than for electrons: $\delta_h\gg\delta_e$. The role of Dresselhaus-type terms
increases with increasing $k_F$. These BIA mechanisms of breaking $s_z$ symmetry
are characterized by the corresponding symmetry breaking rates,
\begin{eqnarray}
\frac{1}{\tau_\Delta}&\sim& \Delta_0^2\tau,
\label{7}\\
\frac{1}{\tau_\delta^{e,h}}&\sim &(\delta_{e,h} k_F)^2\tau,
\label{Dress}
\end{eqnarray}
within the Dyakonov-Perel spin relaxation mechanism (here $\tau$ is the transport scattering
time due to disorder).

The BIA-induced terms are often neglected in the consideration of the quantum spin Hall
effect in view of small coupling constants.
Nevertheless, the break-down of the spin symmetry certainly occurs at sufficiently
long scales (low temperatures), leading in the infrared limit
to a single copy of a symplectic-class system (WAL without doubling).

(ii) Another way to break $s_z$ symmetry is provided by the Rashba term arising
in an
asymmetric well (e.g., due to a finite gate voltage):\cite{Liu08,Rothe10}
\begin{multline}
H_R=\begin{pmatrix}
      0 & 0 & 2 i r_0 k_- & 0 \\
      0 & 0 & 0 & 0 \\
      - 2 i r_0 k_+ & 0 & 0 & 0 \\
      0 & 0 & 0 & 0
    \end{pmatrix}\\
  = r_0(\sigma_0+\sigma_z)(-k_x s_y + k_y s_x).
\label{8}
\end{multline}
Here we keep only the leading (linear-in-$k$) Rashba term connecting E1 up and E1 down subbands;
other terms involving H1 bands are of higher order in $k$.
Within the Dyakonov-Perel mechanism, the corresponding symmetry-breaking rate is
\begin{equation}
\frac{1}{\tau_R}\sim (r_0 k_F)^2\tau,
\label{7R}
\end{equation}

(iii) The diagonal structure of $H_\text{BHZ}$ is broken by atomically sharp
impurities or interface roughness. This type of disorder, combined with the strong
spin-orbit coupling in HgTe, leads to random non-diagonal terms in the Hamiltonian.
Furthermore, the $s_z$ symmetry can also be broken by
finite-range (smooth on the lattice scale) impurities located in the quantum well.
Such impurities break locally the $z\to -z$ symmetry of the QW and can be thought
of as a kind of random Rashba term. Near the transition point, the leading
(linear-in-$k$) contribution comes from the disorder-induced mixing of E1-up and E1-down
blocks, similarly to the Rashba term.
When disorder is created by distant charged impurities, such mixing is suppressed
by an additional factor $d/R$, where $d$ is the QW thickness and $R$ is the spacer width.

\subsection{Symmetry breaking mechanisms at the boundary}
\label{s2.3}

When the dephasing length $L_\phi$ is longer than one of the dimensions of a sample,
the boundary conditions become important for the interference effects.
This is, in particular, the case for universal conductance fluctuations in
a coherent sample. Another example is the conductance of narrow stripes of
HgTe in the regime $L_\phi\gg W$. While the width of the
quantum well is of the order of few nanometers, its lateral width $W$ can be
restricted to about hundred nanometers. In such a restricted quasi-1D geometry,
boundary conditions play a crucial role. Let us analyze the symmetries of the
BHZ Hamiltonian with the boundary conditions.

We first assume that the boundary does not violate the spin symmetry
and the Hamiltonian decomposes into two time reversed blocks $h(\mathbf{k})$ and
$h^*(-\mathbf{k})$. Then the block-wise time reversal symmetry of
$h(\mathbf{k})$ will be inevitably broken at the boundary. The easiest way to
see this is to consider the Dirac limit of the Hamiltonian neglecting the mass
in Eq.\ (\ref{2}). Schr\"odinger equation with such a linear-in-$k$ Hamiltonian
is a system of two coupled linear first order differential equations. Boundary
condition for such a system should be of the zero order in momentum, i.e.,
some linear constraint on the components of the wave function is imposed at the
boundary. In a general form, we can write $b^T \psi = 0$ with some two component
spinor $b$. The elements of $b$ may depend on the conserved momentum parallel to
the boundary. The time-reversal symmetry tells us that, if $\psi$ is an
eigenstate, then $s_y \psi^*$ is another eigenstate with the same energy. If
boundary conditions preserve this symmetry then the latter eigenfunction would
obey $b^T s_y \psi^* = 0$. Equivalently, $b^\dagger s_y \psi = 0$. Thus we have
two linear conditions on $\psi$ instead of one. These conditions are consistent
only if the vectors $b$ and $s_y b^*$ are linearly related: $b = a s_y b^*$ with
some constant $a$. Multiplying the last expression by $b^\dagger$ from the left,
we obtain $b^\dagger b = a b^\dagger s_y b^* = a \mathop{\mathrm{tr}} (s_y b^*
b^\dagger)$. The left-hand side is strictly positive while the right-hand side
contains a trace of the product of antisymmetric ($s_y$) and symmetric
($b^\dagger b^*$) matrices and is hence zero. This apparent controversy proves
that any boundary conditions for a single copy of massless Dirac Hamiltonian will inevitably
break its time-reversal invariance.

The simplest way to see this is as follows:
in order to produce a ``wall'' for a single Dirac-fermion species,
one has to open a gap by switching
on a big mass near the boundary, which breaks the effective
time reversal symmetry already after a single boundary scattering event.
This fact is well known in the context of graphene studies, where in the absence
of the intervalley scattering it is impossible to confine Dirac quasiparticles
without opening the gap at the boundary.

An alternative boundary condition $\psi = 0$ for the BHZ Hamiltonian is widely
used in literature. In order to apply this boundary condition, the quadratic
terms in the Hamiltonian should be retained. The relation $\psi = 0$ is
invariant under time-reversal transformation. Nevertheless, the time-reversal
symmetry is broken at the boundary even in this case. A detailed proof of the
symmetry breaking is relegated to Appendix \ref{App:boundary}.

We thus conclude that  the block-wise symplectic time-reversal
symmetry is always broken near the boundary. 
In particular, in a quasi-1D system the corresponding 
symmetry-breaking rate is given by
\begin{equation}
 \frac{1}{\tau_\text{edge}}
  \sim \frac{D}{W^2},
\end{equation}
where $W$ is the stripe width. This $\tau_\text{edge}$ is just the average time
for an electron to diffuse across the system and get aware of the boundary
conditions.

When the block-wise time-reversal symmetry is of the orthogonal type
(e.g., when the Fermi energy is located near the bottom of the spectrum),
the boundary condition $\psi = 0$ does not introduce additional symmetry breaking,
in contrast to the ``Dirac case''. Indeed, the system belonging to the
conventional orthogonal class can be confined by large potential. Then 
the ``relativistic corrections'' would interplay with the boundary scattering
just in the same way as with the impurity scattering. 

So far, we have considered the boundary conditions preserving the spin symmetry.
The spin symmetry may be
broken by the edges of a 2D sample if the edges are not ideal in $z$-direction.
This situation, which can be modeled by short-range impurities located near the
boundaries,
seems to be quite likely in a realistic setup.
In this case the only remaining symmetry is the physical (symplectic)
time-reversal symmetry.

\section{Interference corrections from symmetry consideration}
\label{s3}

In this section we describe the general symmetry-based formalism
for calculating the interference corrections in an infinite 2D sample.
We closely follow the approach~\cite{Ostrovsky06} developed for Dirac fermions in a
disordered graphene (see also Ref. \onlinecite{McCann}).

Quite generally, the existence of a singular (logarithmic in $T$ or $B$) conductivity correction
is related to the presence of a certain time-reversal (TR) symmetry that acts on an operator 
$\mathcal{O}$ according to
\begin{equation}
T: \quad \mathcal{O}
  \mapsto U^{-1} \mathcal{O}^T U,
\end{equation}
where $U$ is some unitary operator (note that
the momentum operator changes sign under transposition).
For the $4\times 4$ matrix Hamiltonian one can
introduce 16 possible TR symmetries
\begin{equation}
T_{ij}: \quad \mathcal{O}
  \mapsto \sigma_i s_j \mathcal{O}^T s_j \sigma_i,
  \label{12}
\end{equation}
that would correspond to 16 soft modes (Cooperons).
For example, the Hamiltonian $H_{2O}=[\epsilon(k)+V(\mathbf{r})]\sigma_0s_0$
(physically, this Hamiltonian corresponds to spinful electrons in two valleys of a
normal metal, without any spin and valley mixing effects)
is invariant under all these 16 symmetries. This gives rise to 16 logarithmic
corrections to the conductivity of the form
\begin{equation}
\delta \sigma_{ij} = - c_{ij} \frac{e^2}{2 \pi h}
\ln\left(\frac{\tau_\phi}{\tau}\right), \qquad c_{ij}=\pm 1.
\label{deltasigmasym}
\end{equation}
The sign factors $c_{ij}$ here are determined by the sign of the time-reversal operator
squared $T^2=\pm 1$. In the present case we have
$T_{ij}^2=(\sigma_i s_j)(\sigma_i s_j)^*$.
In particular,
\begin{eqnarray}
\sigma_0 s_0(\sigma_0 s_0)^*&=&1 \quad  \Longrightarrow \quad T_{00}^2=1,
\label{13}\\
\sigma_y s_0(\sigma_y s_0)^*&=&-1 \quad \Longrightarrow \quad T_{y0}^2=-1,
\label{14}\\
\sigma_y s_y(\sigma_y s_y)^*&=&1
\quad \Longrightarrow \quad T_{yy}^2=1,
\end{eqnarray}
and so on.
One can see that out of 16 TR symmetries only 6
(namely, $T_{0y},\ T_{xy},\ T_{y0},\ T_{yx},\ T_{yz},\ T_{zy}$) contribute with the minus sign,
yielding for $H_{2O}$ the total conductivity correction
\begin{equation}
\delta \sigma=-(-6+10)\frac{e^2}{2 \pi h} \ln\left(\frac{\tau_\phi}{\tau}\right)
=-2\times\frac{e^2}{\pi h} \ln\left(\frac{\tau_\phi}{\tau}\right).
\end{equation}
This is the WL correction for two independent copies of orthogonal symmetry class.

In general, not all the possible TR symmetries are respected by the dominant
term in the Hamiltonian. Therefore, as a first step in calculating the logarithmic
correction one should retain only the TR symmetries of the dominant term,
setting all other $c_{ij}=0$.
The subleading terms in the Hamiltonian may further break the remaining TR
symmetries on the scale $\tau_{sb}$, introducing gaps in the soft modes and thus cutting
off the logarithmic terms by $\tau_{sb}$.
As a result, in the low-$T$ limit, the singular term $\ln(\tau_\phi/\tau)$ remains
only for those TR symmetries that are preserved by all terms in the Hamiltonian.
With increasing $T$, $\tau_\phi$ becomes shorter than the symmetry breaking time
$\tau_{sb}$ and the system crosses over to another symmetry class.

\subsection{Weak antilocalization in HgTe quantum wells: $E_F\gg m(k_F)$}
\label{s3.1}

Let us now return to the case of HgTe and employ this machinery to it.
In order to directly use the results of Ref. \onlinecite{Ostrovsky06},
we interchange E1 and H1 states in the spin-up block. In this basis
(H1+,E1+,E1-,H1-), the BHZ Hamiltonian (\ref{1}) takes the form:
\begin{equation}
H_{BHZ}=\epsilon(\mathbf{k})\sigma_0 s_0
+ [-m(\mathbf{k}) \sigma_z + A \mathbf{k} {\boldsymbol\sigma}] s_z.
\label{9}
\end{equation}
The linear-in-$k$ (massless Dirac) term in this $4\times 4$ Hamiltonian has the same structure
as in Ref. \onlinecite{Ostrovsky06} (there the matrices $\tau_i$
corresponding to two graphene valleys
played a role of $s_i$ here).

We will now consider the situation when this term in the
Hamiltonian is dominant, i.e. when the Fermi energy $E_F\gg m(k_F)$ is
in the range of almost linear spectrum
$E_\pm \simeq \pm A |k|$, so that the mass term as well as the spin-symmetry breaking terms
can be considered as perturbations.
In this representation, the $s_z$-symmetry breaking terms read:
\begin{align}
H_\text{BIA}
  &= \Delta_0 \sigma_z s_x + \delta_+ (k_x\sigma_x+k_y\sigma_y) s_x \\
  &\qquad +  \delta_-(k_x\sigma_y-k_y\sigma_x) s_y,
  \label{10}
  \\
H_R &= r_0[(k_x\sigma_y+k_y\sigma_x)s_x-(k_x\sigma_x-k_y\sigma_y)s_y],
\label{11}
\end{align}
where $\delta_\pm=\delta_h\pm\delta_e.$

The massless Dirac Hamiltonian (``graphene Hamiltonian'')
\begin{equation}
H_A=A(k_x\sigma_x+k_y\sigma_y)s_z
\label{15}
\end{equation}
is invariant under the following four TR symmetries:
\begin{align}
 T_{xx}:&\quad \mathcal{O}
  \mapsto \sigma_x s_x \mathcal{O}^T \sigma_x s_x,\qquad T_{xx}^2=1,
  \label{16a} \\
 T_{y0}:&\quad \mathcal{O}
  \mapsto \sigma_y s_0 \mathcal{O}^T \sigma_y s_0,\qquad T_{y0}^2=-1,
  \label{16b} \\
 T_{yz}:&\quad \mathcal{O}
  \mapsto \sigma_y s_z \mathcal{O}^T \sigma_y s_z,\qquad T_{yz}^2=-1,
  \label{16c} \\
 T_{xy}:&\quad \mathcal{O}
  \mapsto \sigma_x s_y \mathcal{O}^T \sigma_x s_y,\qquad T_{xy}^2=-1,
  \label{16d}
\end{align}
which in Ref. \onlinecite{Ostrovsky06} were denoted as $T_0, T_x, T_y, T_z$,
respectively.
As a result, the conductivity correction for two independent copies
of massless Dirac fermions is positive:
\begin{equation}
\delta\sigma=-(1-3)\frac{e^2}{2 \pi h} \ln\left(\frac{\tau_\phi}{\tau}\right)
=2\times \frac{e^2}{2 \pi h} \ln\left(\frac{\tau_\phi}{\tau}\right),
\label{2WAL}
\end{equation}
which is a doubled WAL (two copies of symplectic class).

Let us now take into account the mass term $H_M=-m(\mathbf{k})\sigma_zs_z$ in $H_\text{BHZ}$.
This term is invariant under $T_{xx}$ and $T_{xy}$ while breaking down $T_{y0}$ and $T_{yz}$.
The corresponding symmetry breaking rate $1/\tau_m$ was calculated in Ref. \onlinecite{Tkachov11}.
The conductivity correction in this case can be written as
\begin{multline}
 \delta\sigma
  = \frac{e^2}{2 \pi h} \Bigg[
       \ln \frac{\tau}{\tau_\phi}
      -\ln\left( \frac{\tau}{\tau_\phi}+\frac{\tau}{\tau_m}  \right) \\
      -\ln\left( \frac{\tau}{\tau_\phi}+\frac{\tau}{\tau_m}  \right)
      -\ln \frac{\tau}{\tau_\phi}
    \Bigg] \\
  = -2\times \frac{e^2}{2 \pi h}\ln\left(
      \frac{\tau}{\tau_\phi}+\frac{\tau}{\tau_m}  \right).
\end{multline}
At lowest $T$, when $\tau_\phi\gg\tau_m$, we have
\begin{equation}
 \delta\sigma
  \simeq 2\times \frac{e^2}{2 \pi h}\ln\left(\frac{\tau_m}{\tau}  \right),
  \label{2U}
\end{equation}
which is not singular in $T$. This corresponds to two copies of a unitary class.
With increasing $T$ the symmetry-breaking term $1/\tau_m$ becomes unimportant and
the system crosses over to two copies of symplectic class, Eq. (\ref{2WAL}).

So far, we have considered the block-diagonal $s_z$-symmetric Hamiltonian $H_\text{BHZ}$.
Let us now take into account the inversion-asymmetry terms $H_\text{BIA}$ and $H_R$.
The main difference between HgTe and graphene is the structure of such symmetry breaking terms.
The intervalley disorder scattering in graphene preserves only $T_{xx}$ symmetry. As a result,
when the valley mixing is strong, the quantum correction to the conductivity is negative (WL):
graphene with mixed valleys belongs to the orthogonal symmetry class.
In contrast, the mixing of the up/down blocks in HgTe QWs involves an additional spin structure
which generically kills the orthogonal $T_{xx}$ symmetry. Generically, the only remaining TR
symmetry is the symplectic $T_{xy}$ symmetry surviving the SO interaction.
Therefore, at lowest temperatures the HgTe QW is expected to be a single copy
of a symplectic system with ordinary WAL.

The $k$-independent term $\Delta_0 \sigma_z s_x$ in $H_\text{BIA}$ violates $T_{xx}$ and
$T_{y0}$, while preserving $T_{yz}$ and $T_{xy}$. This leads to
\begin{multline}
 \delta\sigma
  = \frac{e^2}{2 \pi h} \Bigg[
  \ln\left( \frac{\tau}{\tau_\phi} + \frac{\tau}{\tau_\Delta} \right)
  -\ln\left(
        \frac{\tau}{\tau_\phi} + \frac{\tau}{\tau_m} + \frac{\tau}{\tau_\Delta}
      \right) \\
            -\ln\left( \frac{\tau}{\tau_\phi} + \frac{\tau}{\tau_m} \right)
            -\ln\frac{\tau}{\tau_\phi}
    \Bigg].
    \label{sigmaDelta}
\end{multline}
Two situations are possible: (i) $\tau_\Delta\gg\tau_m$ and (ii) $\tau_\Delta\ll\tau_m$.
Starting at high $T$, with lowering $T$ in case (i) the system first crosses over from two
copies of symplectic class, Eq. (\ref{2WAL}), to two copies of unitary class, Eq. (\ref{2U}),
and then at $\tau_\phi\sim\tau_\Delta$ to
a single copy of a symplectic class with ordinary WAL:
\begin{equation}
\delta\sigma=\frac{e^2}{2 \pi h} \ln\left(\frac{\tau_\phi}{\tau}\right).
\label{1WAL}
\end{equation}
We denote such an evolution as 2Sp $\to$ 2U $\to$ 1Sp.
In case (ii), for $\tau_\Delta\ll\tau_\phi$ we get
\begin{equation}
 \delta\sigma
  \simeq \frac{e^2}{2 \pi h} \left[
              -\ln\left( \frac{\tau}{\tau_\phi} + \frac{\tau}{\tau_m} \right)
            -\ln\frac{\tau}{\tau_\phi}
    \right].
\end{equation}
which corresponds to the crossover 2Sp $\to$ 1Sp at $\tau_\phi\sim\tau_m$ (2WAL $\to$ 1WAL).
Note that at $\tau_\phi\sim\tau_\Delta$ the number of soft modes does not change, despite the
transitions between the spin blocks.

Finally, in the presence of Dresselhaus and/or Rashba terms, the only effective TR symmetry is
$T_{xy}$. The effect of short-range impurities within the QW is similar to that of the Rashba
term. Denoting the total symmetry-breaking rate due to such terms as $1/\tau_{SO}$,
we get (here neglecting the $\Delta_0$-term):
\begin{multline}
 \delta\sigma
  = \frac{e^2}{2 \pi h} \Bigg[
  \ln\left( \frac{\tau}{\tau_\phi} + \frac{\tau}{\tau_\text{SO}} \right) \\
  -2\ln\left(
        \frac{\tau}{\tau_\phi} + \frac{\tau}{\tau_m} + \frac{\tau}{\tau_\text{SO}}
      \right)
            -\ln\frac{\tau}{\tau_\phi}
    \Bigg].
    \label{sigmaSO}
\end{multline}
Again, for $\tau_m\ll\tau_\text{SO}$, we have 2Sp $\to$ 2U $\to$ 1Sp,
whereas for $\tau_m\gg\tau_\text{SO}$
the mass term is not important (2Sp $\to$ 1Sp).

The general formula for the conductivity correction at $E_F\gg m(k_F)$
including all symmetry-breaking terms combines Eqs. (\ref{sigmaDelta}) and (\ref{sigmaSO}):
\begin{multline}
 \delta\sigma
  = \frac{e^2}{2 \pi h} \Bigg[
  \ln\left( \frac{\tau}{\tau_\phi} + \frac{\tau}{\tau_\Delta}  +
\frac{\tau}{\tau_\text{SO}} \right) \\
  -\ln\left(
        \frac{\tau}{\tau_\phi} + \frac{\tau}{\tau_m} +
        \frac{\tau}{\tau_\Delta} + \frac{\tau}{\tau_\text{SO}}
      \right) \\
            -\ln\left( \frac{\tau}{\tau_\phi} + \frac{\tau}{\tau_m} +
              \frac{\tau}{\tau_\text{SO}} \right)
            -\ln\frac{\tau}{\tau_\phi}
    \Bigg].
    \label{SpFull}
\end{multline}
Here $1/\tau_\phi$ is the dephasing rate due to inelastic processes,
$\tau$ is elastic transport scattering time,
$\tau_\Delta$ was defined in Eq.~(\ref{7}),
$1/\tau_{SO}$ is the total spin-orbit rate describing the combined effect of Dresselhaus and Rashba
terms, Eqs. (\ref{Dress}) and (\ref{7R}) (as well as the SO impurity
scattering), $1/\tau_m$ is the rate of breaking effective
time-reversal symmetry
within each spin block due to finite mass of Dirac fermions.
This mechanism of symmetry breaking is analogous to the one in the
case of interplay between Rashba and Zeeman splitting:
the Rashba term tends to fix the spin direction in the 2D plane according to the momentum of a particle,
whereas the Zeeman field tends to align the spin in the direction of magnetic field (out of the 2D plane).
Without impurity scattering, each state is characterized by a certain direction of spin.
The symmetry-breaking rate can be estimated using an analogy with the Dyakonov-Perel mechanism,
$\tau_m^{-1}\sim \Delta^2_m \tau$, where effective spin-precession frequency
$\Delta_m$ is only non-zero in the presence of disorder:
\begin{equation}
\Delta_m \propto \frac{m(k_F)}{Ak_F} \frac{1}{\tau},
\end{equation}
which yields
\begin{equation}
\frac{1}{\tau_m}\sim \frac{1}{\tau}\, \left[\frac{m(k_F)}{E_F}\right]^2
\end{equation}
in the regime of interest.
A rigorous calculation of a Cooperon \cite{kachor} (which can be simplified by dressing the
impurity potential by Dirac factors, thus reducing the problem to a spinless one in the
presence of a peculiar disorder potential) confirms this estimate.
This result agrees with
Ref. \onlinecite{Tkachov11} in this limit.

Assuming for simplicity an energy-independent disorder scattering rate 
(which, in fact, is generically not 
the case for Dirac particles, see, e.g., Ref. \onlinecite{Ostrovsky06}),
we can express the symmetry-breaking length $l_m=(D\tau_m)^{1/2}$
in terms of the carrier density $n$:
\begin{equation}
 l_m \propto \dfrac{A}{\left|\pm \dfrac{|M|}{\sqrt{n}}+B\sqrt{n}\right|},
\label{lmn}
\end{equation}
where the sign $\pm$ distinguishes between the inverted ($-$) and normal
($+$) band structures.

We have identified the only true soft mode (corresponding to the
true Kramers TR symmetry), which
governs the localization properties at lowest temperatures. It gives
rise to a single WAL correction (single symplectic class: 1Sp).
This is not surprising, as the symplectic TR symmetry is the only true TR symmetry in a system
with strong spin-orbit interaction.
On the other hand, on short time scales (equivalently, high
temperatures) the symmetry may be higher if the corresponding
symmetry-breaking terms are not yet operative. Depending on the values of symmetry-breaking rates,
there may be different patterns of symmetry breaking:
\begin{itemize}
\item[(i)] 2Sp $\to$ 2U $\to$ 1Sp. This first scenario is realized
when the mass term in $H_\text{BHZ}$ is more important than the $s_z$-symmetry-breaking terms:
 $\tau\ll\tau_m\ll\text{min}[\tau_\Delta,\tau_{SO}]$. In this case, at
 high temperatures the system behaves as two
  copies of symplectic class (2Sp) with doubled WAL. In the intermediate range of temperatures
 $\tau_m\ll\tau_\phi\ll\text{min}[\tau_\Delta,\tau_{SO}]$, the system
 behaves as two copies of unitary class (2U)
 with no $T$-dependent interference correction. Finally, at
 $\tau_\phi\gg\text{min}[\tau_\Delta,\tau_{SO}]$
 the two spin blocks are completely mixed and the system reaches its
 generic spin-orbit state 1Sp (single copy
 of symplectic class with ordinary WAL).
\item[(ii)] 2Sp $\to$ 1Sp. This scenario is realized in the case of
  the following hierarchy of scales:
$\tau\ll\text{min}[\tau_\Delta,\tau_{SO}]\ll\tau_m$, when the block
mixing is faster than the breaking of the
effective TR symmetry within each block. As a result, there is no room
for the unitary class in this case.
\item[(iii)] 1Sp. When the complete mixing of spin blocks is very fast, which happens when
$\tau_\text{SO}\alt \tau$ or $\text{max}[\tau_m,\tau_\Delta]\alt\tau$,
the system behaves as a single copy of symplectic
class (1Sp) in the whole diffusive regime (ordinary WAL).
\end{itemize}

Let us estimate which of these scenarios can be expected in experiments on HgTe QWs.
According to Table I of Ref. \onlinecite{Qi11}, the value of the BIA-induced splitting $\Delta_0$
in HgTe QW near the QSHE transition is $\Delta_0\simeq 0.0015 - 0.002 \text{eV} \sim 15 - 20 K$.
When the 2D mean free path is $0.1-0.5 \mu$m, one finds (using $A \simeq 3.65 - 3.9$ eV \AA)
\begin{equation}
\Delta_0\tau \simeq 0.5 - 2.5 \sim 1.
\end{equation}
This implies that for such values of the mean free path there is no room for scenario (i).
The conductivity correction for $\Delta_0 \tau \agt 1$ has the form
 \begin{equation}
 \delta\sigma
  \simeq \frac{e^2}{2 \pi h} \left[
              -\ln\left( \frac{\tau}{\tau_\phi} + \frac{\tau}{\tau_m}+\frac{\tau}{\tau_{SO}} \right)
            -\ln\frac{\tau}{\tau_\phi}
    \right].
    \label{2-1}
\end{equation}
Then for $\text{min}[\tau_m,\tau_\text{SO}]\gg \tau$ the pattern (ii) 2Sp $\to$ 1Sp
is realized, while for $\text{min}[\tau_m,\tau_\text{SO}]\alt \tau$
there is single copy of symplectic system (1Sp)
in the diffusive regime.

In a perpendicular magnetic field, a magnetoresistance arises due to
the suppression of the interference
correction by magnetic field. In the diffusive regime $L_\phi, L_H \gg
l$ [where $L_H=(\hbar c/e B)^{1/2}$ is
the magnetic length and $l$ is the mean free path],
the effect of magnetic field can be described by the replacement
\begin{equation}
\frac{1}{\tau_\phi}\to \frac{1}{\tau_\phi}+\frac{D}{L_H^2}
\label{tauH}
\end{equation}
in the above zero-$B$ formulas (here $D$ is the diffusion coefficient).
A more accurate expression (valid also in the limit of weak magnetic field, $D\tau_\phi/L_H^2\ll 1$)
has the standard Hikami-Larkin-Nagaoka \cite{hln} form with digamma functions replacing logarithms.
For a single copy of symplectic system, Eq.~(\ref{1WAL}), the magnetoconductivity $\Delta\sigma(B)=\sigma(B)-\sigma(0)$
has the from
\begin{eqnarray}
\Delta\sigma(B)&=&- \frac{e^2}{2\pi h}\,
{\cal H}\left(\frac{\tau}{\tau_\phi},\frac{B}{B_{tr}}\right), \nonumber \\
{\cal H}(x,y)& = & \psi\left(\frac{1}{2}+\frac{x}{y}\right) -
\psi\left(\frac{1}{2}+\frac{1}{y}\right)- \ln{x}, \label{eq43}
\end{eqnarray}
where $B_{tr}=\hbar/(2el_{tr}^2)$ is the value of magnetic field for which $L_H$ is equal
to the transport mean-free path $l_{tr}$ and $\psi(x)$ is a digamma function.
When the zero-$B$ expression contains a combination of several logarithms representing the contributions
of various channels, 
in the formula for the magnetoconductivity, each of logarithms is to be replaced according to Eq.~(\ref{eq43}),
with $1/\tau_\phi \to 1/\tau_\phi+1/\tau_{sb}$, where $1/\tau_{sb}$ is the total symmetry 
breaking rate for given channel.
Weak antilocalization leads then to a positive magnetoresistance.

Importantly, in the regime $E_F\gg m(k_F)$ considered above there is no room for
weak localization behavior which would manifest itself as a negative magnetoresistance.
However, in Refs. \onlinecite{kvon,minkov} a crossover form positive to negative magnetoresistance
was observed with increasing magnetic field in samples with not too high carrier densities.
Such a crossover is characteristic to
conventional spin-orbit systems with $\tau_{SO}\gg \tau$.
In order to describe this behavior in HgTe quantum wells we have to consider the opposite limit $E_F\ll m(k_F)$,
where the off-diagonal (``spin-orbit'') terms in the Hamiltonian can be considered as a small
perturbation.

\subsection{Weak localization in HgTe quantum wells: $E_F\ll m(k_F)$}
\label{s3.2}

Let us now consider the limit of low densities, when the Fermi energy
is smaller than $m$. This is the limit of the conventional Schr\"odinger equation
in the relativistic Dirac description. In this case, the spectrum can be expanded in $A^2k^2$ (for definiteness, we consider the upper branch):
\begin{eqnarray}
 E_+(\mathbf{k})
  &=& \epsilon(\mathbf{k}) + \sqrt{A^2 k^2 + m^2(\mathbf{k})} 
\nonumber \\
&\simeq& |M| + B k^2 +
    \epsilon(\mathbf{k})+\frac{A^2 k^2}{|M|}.
  \label{spectrum}
\end{eqnarray}
In this case the pseudospin is almost fixed along the $z$-direction by the ``Zeeman field'' $M$, while the off-diagonal terms in the Hamiltonian
can be viewed as a weak ``relativistic'' correction (analogous to spin-orbit
coupling), see Appendix \ref{App:bottom}. On short
scales the system is therefore close to the
orthogonal symmetry class and should exhibit negative magnetoresistance in sufficiently strong magnetic field.

As discussed above, a similar situation is possible for the BHZ Hamiltonian at very high densities, when $B k_F^2 \gg \{|M|, A k_F\}$,
where the spectrum is given by
\begin{equation}
E_+(\mathbf{k})
   \simeq B k^2 + \epsilon(\mathbf{k})+M+\frac{A^2}{B}.
\end{equation}
However, in realistic systems, when the first term in the small-$k$ expansion becomes dominant, the expansion is expected
to break down: all higher-order terms are expected to be as relevant as the lowest-order one ($Bk^2$ here).
The analysis of the system at such high energies requires a more detailed knowledge of the spectrum.
Therefore, here we will focus on the controllable case of small energies.

Now we consider the massless Dirac Hamiltonian
$
H_A=A(k_x\sigma_x+k_y\sigma_y)s_z
$
as a perturbation to the diagonal part of the BHZ Hamiltonian
\begin{equation}
H_M= - m(\mathbf{k}) \sigma_z  s_z,
\label{HamMass}
\end{equation}
dominated by the mass term $m(\mathbf{k}) = M + B k^2$.
The Hamiltonian (\ref{HamMass})
is invariant under the following eight TR symmetries:
\begin{align}
T_{00}:&\quad \mathcal{O}
  \mapsto \sigma_0 s_0 \mathcal{O}^T \sigma_0 s_0,\qquad T_{00}^2=1,  \label{Tmass00}\\
T_{0z}:&\quad \mathcal{O}
  \mapsto \sigma_0 s_z \mathcal{O}^T \sigma_0 s_z,\qquad T_{0z}^2=1,  \label{Tmass0z}\\
T_{xx}:&\quad \mathcal{O}
  \mapsto \sigma_x s_x \mathcal{O}^T \sigma_x s_x,\qquad T_{xx}^2=1,  \label{Tmassxx}\\
T_{xy}:&\quad \mathcal{O}
  \mapsto \sigma_x s_y \mathcal{O}^T \sigma_x s_y,\qquad T_{xy}^2=-1,  \label{Tmassxy}\\
T_{yx}:&\quad \mathcal{O}
  \mapsto \sigma_y s_x \mathcal{O}^T \sigma_y s_x,\qquad T_{yx}^2=-1, \label{Tmassyx}\\
T_{yy}:&\quad \mathcal{O}
  \mapsto \sigma_y s_y \mathcal{O}^T \sigma_y s_y,\qquad T_{yy}^2=1, \label{Tmassyy}\\
T_{z0}:&\quad \mathcal{O}
  \mapsto \sigma_z s_0 \mathcal{O}^T \sigma_z s_0,\qquad T_{z0}^2=1,  \label{Tmassz0}\\
T_{zz}:&\quad \mathcal{O}
  \mapsto \sigma_z s_z \mathcal{O}^T \sigma_z s_z,\qquad T_{zz}^2=1.  \label{Tmasszz}
\end{align}
We see that there are overall six symmetries of the ``orthogonal'' type
(yielding WL) and two ``symplectic'' (yielding WAL). However, we have to take into account that the Hamiltonian
$H_M$ breaks into blocks corresponding to positive and negative energies
(conduction band and valence band). The two blocks correspond to eigenvalues of
$\sigma_z s_z$ equal to $\pm 1$. The transport is determined only by one of
these blocks (where the Fermi energy is located).

Therefore, eight symmetries Eqs.\ (\ref{Tmass00}) -- (\ref{Tmasszz}) split in
four pairs,
\begin{equation}
\begin{aligned}
 T_{00} &\sim T_{zz}, \\ 
 T_{0z} &\sim T_{z0}, \\
 T_{xx} &\sim T_{yy}, \\ 
 T_{xy} &\sim T_{yx},
 \end{aligned}
 \label{pairs}
\end{equation}
and each pair should be counted only once for the calculation of the WL
correction since two symmetries of the pair become identical when reduced to
any of two $\sigma_z s_z$ eigenblocks.
This yields
\begin{equation}
\delta \sigma=-(3-1)\frac{e^2}{2 \pi h} \ln\left(\frac{\tau_\phi}{\tau}\right)
=-\frac{e^2}{\pi h} \ln\left(\frac{\tau_\phi}{\tau}\right),
\label{2O}
\end{equation}
that is the contribution of an antilocalizing singlet and a localizing triplet.
This is a conductivity correction for two independent copies of orthogonal
symmetry class.

Let us now include the off-diagonal Dirac term $H_A =
A(k_x\sigma_x+k_y\sigma_y)s_z$. The Hamiltonian $H_M + H_A$ still separates
into two blocks with approximate orthogonal symmetry but positive- and
negative-energy sectors are now mixed. Effective Hamiltonian valid near the
Fermi energy is derived in Appendix \ref{App:bottom}. Breaking of the
time-reversal symmetry within a single block of the BHZ Hamiltonian can be
traced back to the appearance of an effective magnetic field [with the vector
potential given by Eq.\ (\ref{vector})] due to the interplay of ``relativistic
corrections'' and  disorder scattering. The weak localization correction
(\ref{2O}) is modified by the presence of this effective magnetic field,
\begin{equation}
\delta \sigma=\frac{e^2}{\pi h}
\ln\left(\frac{\tau}{\tau_A}+\frac{\tau}{\tau_\phi}\right),
\label{2O2U}
\end{equation}
where $1/\tau_A$ is the symmetry-breaking rate due to $H_A$ (it is analogous
to $1/\tau_M$ introduced above). The value of $1/\tau_A$ can be estimated as
follows.
Adding disorder potential to $H_\text{BHZ}$, we project the full Hamiltonian on
a single chiral branch. The off-diagonal part $H_A$ would lead to the appearance
of the
``spin-orbit'' impurity scattering characterized by
\begin{equation}
\Delta_A \sim \frac{1}{\tau} \frac{A^2 k_F^2}{m^2(k_F)}.
\end{equation}
This estimate yields 
\begin{equation}
\frac{1}{\tau_A} \sim \Delta_A^2 \tau \sim \frac{1}{\tau} \left[\frac{A k_F}{m(k_F)}\right]^4.
\label{tauA}
\end{equation}

Alternatively, using the reduced spinless Hamiltonian (\ref{heff}), one can 
evaluate the averaged phase difference $S_\text{a}$ (``dephasing action'') accumulated due to the effective vector 
potential, Eq.~(\ref{vector}), on the time-reversed paths within the path-integral formalism
(for definiteness, we consider the case when the Fermi level  is located near 
the bottom of the spectrum, $E_F\ll |M|$):
\begin{eqnarray}
S_\text{a} &\sim& e^2 \left\langle \int d\mathbf{l}_1 \mathbf{a}(\mathbf{r}_1) 
\int d\mathbf{l}_2 \mathbf{a}(\mathbf{r}_2)\right\rangle_\text{dis}\nonumber \\
&\sim& \frac{1}{M^2} \left\langle \int d\mathbf{l}_1 \times \nabla V(\mathbf{r}_1)
\int d\mathbf{l}_2 \times \nabla V(\mathbf{r}_2)\right\rangle_\text{dis}.\nonumber \\
\end{eqnarray}
Using the short-range correlated potential,
$$\langle V(\mathbf{r}_1) V(\mathbf{r}_2)\rangle_\text{dis} 
= \frac{\delta(\mathbf{r}_1-\mathbf{r}_2)}{2\pi \nu \tau},$$
one can estimate
\begin{equation}
 S_\text{a} \sim \frac{v_F k_F^3}{M^2 \nu \tau}\, t,
\label{Saresult}
\end{equation}
where 
$\nu \sim |M|/A^2$ is the density of states,
$v_F=k_F A^2/|M|$ is the Fermi velocity, 
and all the gradients, as well as the inverse size of impurity, 
are replaced by $k_F$.
From $S_\text{a} \sim 1$ we find $t\sim M^2\nu \tau/v_F k_F^3$,
which yields Eq. (\ref{tauA}).
Again, the derivation~\cite{kachor} of a Cooperon within the kinetic equation
approach and diagrammatics confirms this estimate.

The scaling of the corresponding symmetry-breaking length $l_A=(D \tau_A)^{1/2}$ with the 
carrier concentration $n$ for energy-independent $\tau$ has the form:
\begin{equation}
 l_A \propto \left(\pm \frac{|M|}{\sqrt{n}}+B\sqrt{n}\right)^2.
\label{lAn}
\end{equation}
It is interesting to note that there is a formal relation between 
the symmetry-breaking lengths in the symplectic [Eq.~(\ref{lmn})] and orthogonal [Eq.~(\ref{lAn})] cases:
\begin{equation}
 \frac{l_m}{l}=\sqrt{\frac{l}{l_A}}.
\end{equation}
In each of the cases only one length is relevant,
whereas the other is then shorter than the mean free path,
implying no diffusive dynamics in the corresponding channel.
Both lengths become of the order of the mean free path 
for $E_F \sim m(k_F)$, where without valley mixing the logarithmic
first-order interference correction does not develop at all (2U phase).

Let us now include BHZ block-mixing rates $\tau_\Delta^{-1}$ and
$\tau_{SO}^{-1}$. As we have pointed out above, the eight symmetries
(\ref{Tmass00}) -- (\ref{Tmasszz}) of the Hamiltonian $H_M$ split into four pairs
(\ref{pairs}). The symmetries within each pair are indistinguishable near the
Fermi energy. The general rule for the evaluation of the WL correction is as
follows. For each of the pairs (\ref{pairs}), one of the following three
situations is possible: (i) Both symmetries from the pair are preserved (which
means that the system splits into two eigenblocks of $\sigma_z s_z$). Then the
pair contribute to the WL correction as a single time-reversal symmetry. (ii)
One of two symmetries is broken, the other one is preserved. Then the
situation becomes conventional, and the remaining mode gives a
usual logarithmic contribution. (iii) Both symmetries from the pair are broken.
Clearly, there is no contribution from this pair to the WL (WAL) correction in
this case. To summarize, the pair gives a contribution as a single Cooperon
mode, unless both symmetries are broken. The fact that one should break both
symmetries to suppress the contribution implies that one should add
symmetry-breaking times (rather than rates) corresponding to both
symmetries of a pair and then invert the result in order to get the mass of the
corresponding mode.

In our case additional terms have the following symmetries. The Dirac
kinetic term $H_A$ preserves only $T_{xx}$ and $T_{xy}$ out of the list
(\ref{Tmass00})--(\ref{Tmasszz}). The $\Delta$-term $H_\Delta = \Delta_0
\sigma_z s_x$ conserves $T_{00}$, $T_{z0}$, $T_{xy}$, and $T_{yy}$. The Rashba
and Dresselhaus terms (linear in momentum) conserve $T_{xy}$ and $T_{yx}$.
This yields the following result for the WL correction:\cite{footnoteWL}
\begin{multline}
 \delta\sigma
   = \frac{e^2}{2 \pi h}\Bigg[2 \ln\left(
       \frac{\tau}{\tau_\phi}+\frac{\tau}{\tau_A} +\frac{\tau}{\tau_{SO}}
     \right) \\
     +\ln\left( \frac{\tau}{\tau_\phi}+\frac{\tau}{\tau_{SO}}
     +\frac{\tau}{\tau_A +\tau_\Delta}\right)
     -\ln \frac{\tau}{\tau_\phi} \Bigg].
\end{multline}

\begin{figure*}
 \includegraphics[width=0.51\textwidth]{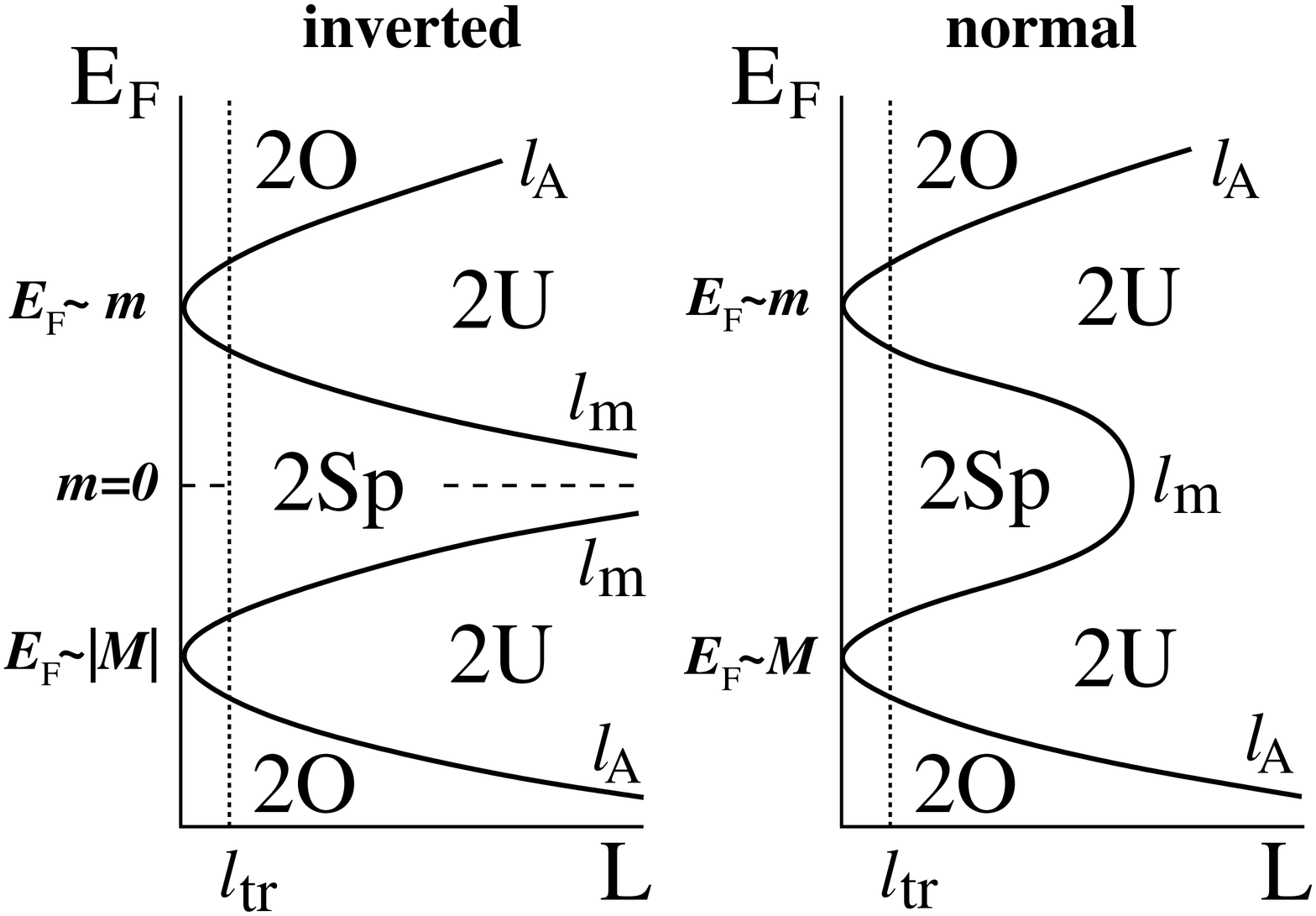}\hspace*{0.2cm} 
\includegraphics[width=0.43\textwidth]{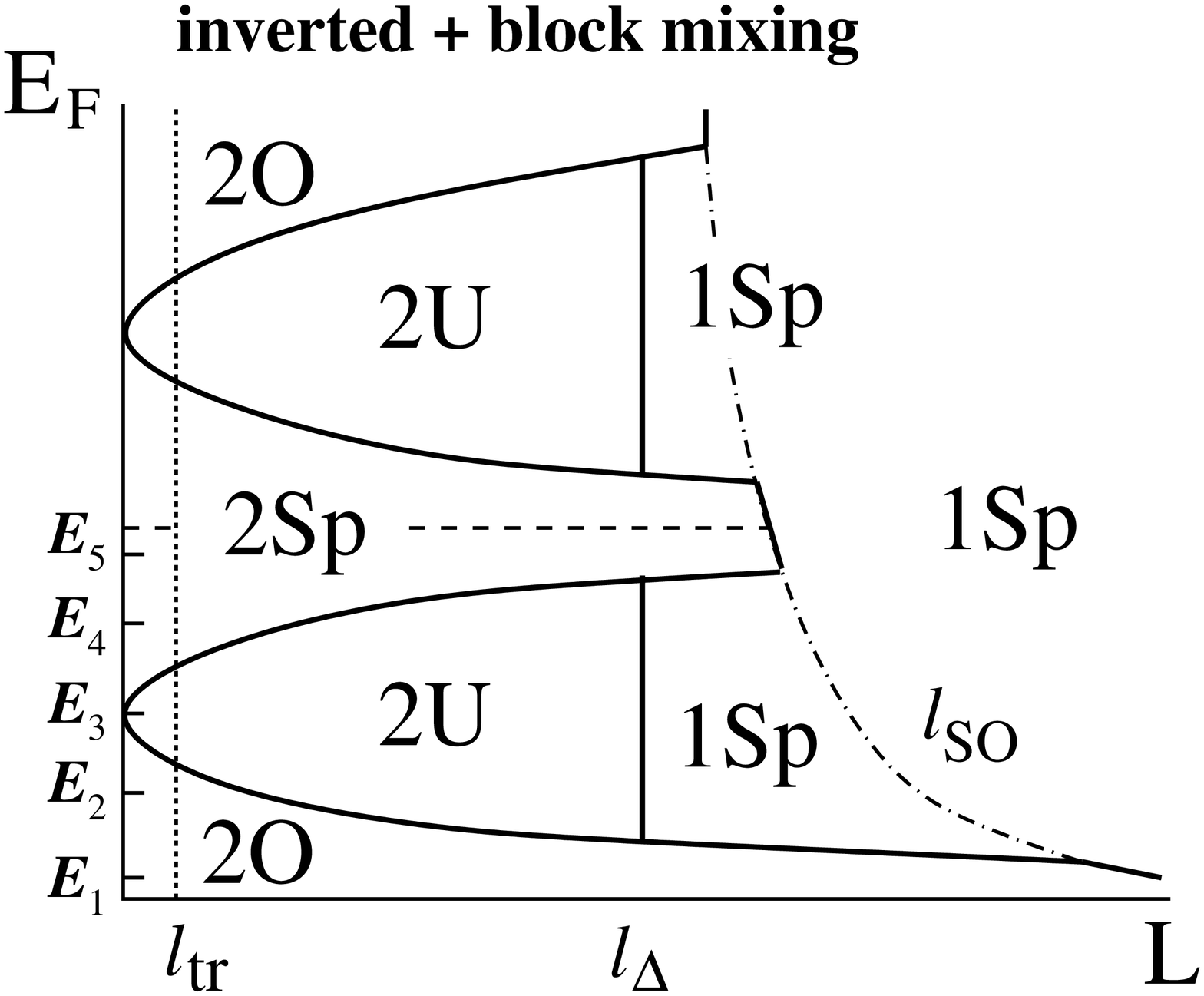} 
 \caption{Schematic diagram showing the symmetry patterns in the $L-E_F$ 
plane. 
 Here $L$ is the system size, phase-breaking length $L_\phi$, or magnetic 
length $L_H$,
 whichever is shorter.  For simplicity, the transport scattering length 
$l_\text{tr}$ (vertical dotted line) is assumed 
 to be energy independent (this assumption does not affect the ``topology'' of 
the diagram).
 The ``phase boundaries'' are shown by solid curves. 
Boundaries of the 2U-regions are determined by 
$l_m=(D\tau_{m})^{1/2}$ 
 for 2Sp $\to$ 2U crossover, and by $l_A=(D\tau_{A})^{1/2}$  for 2O $\to$ 2U 
crossover.
 \textit{Left panel}: Inverted band structure (thick quantum well) with no 
block mixing. Dashed line shows energy for which $m(k_F)=-|M|+Bk_F^2=0.$
 \textit{Middle panel}: Normal band structure (thin quantum well) with no block 
mixing. 
 \textit{Right panel}: Inverted band structure with block mixing, 
 characterized by $l_\Delta=(D\tau_\Delta)^{1/2}$ and 
$l_\text{SO}=(D\tau_{SO})^{1/2}$ (dash-dotted),
assuming $\tau_\Delta<\tau_{SO}$. 
The energies $E_F=E_1\ldots E_5$ (from bottom to top) mark different horizontal 
cross-sections
of the ``phase-diagram'' corresponding to the patterns 2O $\to$ 1Sp, 2O $\to$ 
2U $\to$ 1Sp,
2U $\to$ 1Sp, 2Sp $\to$ 2U $\to$ 1Sp, and 2Sp $\to$ 1Sp, respectively,
that appear with increasing $L$. The magnetoresistance curves corresponding to 
these energies are
shown schematically in Fig. \ref{Fig2}.
The diagram for the case of normal band structure with the block mixing
is qualitatively similar.}\label{Fig1}
\end{figure*}

\begin{figure*}
 \includegraphics[width=0.95\textwidth]{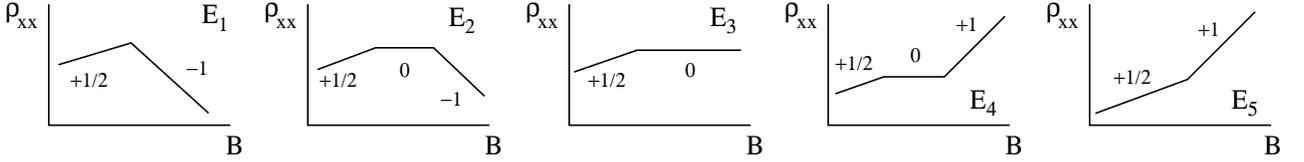}\hspace*{0.2cm}
 \caption{Schematic view of the magnetoresistance curves $\rho_{xx}(B)$ 
 in linear-logarithmic scale. Panels correspond to energies 
$E_1 \ldots E_5$ from the right panel of Fig. \ref{Fig1} (inverted band 
structure).
The numbers indicate the prefactor $\alpha=(\pi h/e^2) \partial \sigma/\partial \tau_H$ (here $\tau_H=L_H^2/D$)
of the logarithmic magnetoresistance in different domains of magnetic field. 
For the purpose of visualization, the crossover regions (where $\tau_H$ is of the order of the corresponding
symmetry-breaking time) are shown as cusps.
}\label{Fig2}
\end{figure*}

The only massless mode here again corresponds to the $T_{xy}$ symmetry (physical symplectic TR symmetry).
Thus in the presence of all block-mixing terms, the system becomes a single copy
of the symplectic class. Depending on the relation between $\tau_A^{-1}$,
$\tau_\Delta^{-1}$, and $\tau_{SO}^{-1}$, three patterns of symmetry breaking
are possible:
\begin{itemize}
\item[(i)] 2O $\to$ 2U $\to$ 1Sp. This scenario is realized
when $\tau\ll\tau_A\ll\text{min}[\tau_\Delta, \tau_{SO}]$. In this case, at
 high temperatures the system behaves as two
  copies of orthogonal class (2O) with doubled WL. In the intermediate range of temperatures
 $\tau_A\ll\tau_\phi\ll \text{min}[\tau_\Delta, \tau_{SO}]$, the system
 behaves as two copies of unitary class (2U)
 with no $T$-dependent interference correction. Finally, at
 $\tau_\phi\gg \text{min}[\tau_\Delta, \tau_{SO}]$
 the two spin blocks are completely mixed and the system reaches its
 generic spin-orbit state 1Sp (single copy
 of symplectic class with ordinary WAL).
\item[(ii)] 2O $\to$ 1Sp. This scenario is realized when
$\text{min}[\tau_\Delta, \tau_{SO}]\ll \tau_A $ and $\tau\ll \text{min}[\tau_A, \tau_{SO}]$, 
i.e., when the block
mixing is faster than the breaking of the
effective orthogonal TR symmetry within each block.
As a result, there is no room
for the unitary class in this case.
\item[(iii)] 1Sp.  When the complete mixing of spin blocks is very fast, which happens when
$\tau_{SO} \alt \tau$,
the system behaves as a single copy of symplectic
class (1Sp) in the whole diffusive regime (ordinary WAL).
\end{itemize}
In the cases (i) and (ii), the magnetoresistance changes from positive to negative with increasing magnetic field.
In the third case, the magnetoresistance is always positive in the diffusive regime.

\section{Discussion and conclusions}
\label{s4}

To summarize, we have explored symmetries of the Bernevig-Hughes-Zhang
Hamiltonian of 2D charge carriers in HgTe/HgCdTe quantum wells 
in the presence of physically relevant symmetry breaking perturbations. We have
identified regimes of different symmetry in the parameter space and evaluated
the corresponding quantum interference corrections to conductivity. Possible
regimes include 2O, 2U, 2Sp, and 1Sp, with the temperature dependence of the 2D
weak (anti-)localization correction given by $\delta\sigma=\alpha (e^2/\pi h) \ln
\tau_\phi(T)$, where $\alpha = -1, 0, 1$, and $1/2$, respectively.  
These regimes are summarized in Fig. \ref{Fig1}.

Experimentally, the quantum interference---weak localization or
antilocalization---shows up most directly in the transverse-field
magnetoresistance. In particularly interesting situations, a symmetry breaking
pattern then determines a succession of regions of magnetic field with different
signs and/or prefactors of the weak localization magnetoresistance (see Fig. \ref{Fig2}). 

Let us now discuss published experimental data \cite{kvon,minkov} in context of
our findings. In these papers the weak localization was
studied in structures with both normal\cite {kvon} and inverted
\cite{kvon,minkov} band gaps. In both cases the authors observed weak
positive magnetoresistance in low magnetic fields $B$ which could be clearly
attributed to WAL. In higher fields, a crossover to negative magnetoresistance
was observed that could be presumably attributed to weak localization. This
corresponds to the $2\,{\rm O} \to 1\,{\rm Sp}$ symmetry pattern [characteristic
for systems with relatively small carrier concentration, $E_F \ll m(k_F)$] in
our terminology. The coefficient in front of $(e^2/\pi h) \ln \tau_H$ (where $\tau_H=L_H^2/D$) was found
to be consistent with  1/2, as expected for the 1Sp regime.  

More experimental work is clearly needed to explore systematically different
parameter regimes with different types of quantum interference behavior.
We hope that our paper will stimulate such experimental activity and will be
helpful in identification of different regimes and analysis of quantum
interference contributions to conductivity.

When this manuscript was in preparation, we learned about preprints
Ref.\ \onlinecite{Glazman,Richter12} with partly overlapping content.

\section{Acknowledgments}

We are grateful C. Br\"une, M. Dyakonov, A. Germanenko, E. Hankiewicz, V. Kachorovskii, G. Minkov, L. Molenkamp, 
S. Tarasenko, and G. Tkachov for numerous illuminating discussions.
This work was supported by BMBF, DFG-RFBR, DFG-SPP ``Semiconductor spintronics'', and DFG-CFN.
P.M.O. and A.D.M. acknowledge the hospitality of Kavli Institute of Theoretical Physics of University of California,
Santa Barbara, where a part of this work was done.

\appendix

\section{Boundary scattering}
\label{App:boundary}

In this Appendix we consider the properties of the BHZ Hamiltonian with the
boundary condition $\psi = 0$, see Sec. \ref{s2.3}.
 We demonstrate that the symplectic (block-wise)
time-reversal symmetry is strongly violated by this boundary condition.

Consider the upper block of the BHZ Hamiltonian (to minimize the notation we
assume $A = 1$)
\begin{equation}
 h
  = \begin{pmatrix}
      M - B \mathbf{k}^2 & k_x - i k_y \\
      k_x + i k_y & -M + B \mathbf{k}^2
    \end{pmatrix}.
\end{equation}
At energy $E = |\mathbf{k}| = \sqrt{M/B}$, diagonal elements of the Hamiltonian
vanish leading to the emergence of symplectic time-reversal symmetry. There are
two plain wave solutions at this energy with the fixed momentum $k_x$ along $x$
direction:
\begin{gather}
 \psi_\pm
  = \begin{pmatrix}
      1 \\ a_\pm
    \end{pmatrix} e^{i k_x x \pm i \sqrt{E^2 - k_x^2} y},
 \\
 a_\pm
  = \frac{k_x \pm i \sqrt{E^2 - k_x^2}}{E}.
\end{gather}
These two states are related by the time-reversal operation:
\begin{equation}
 \sigma_y \psi_\pm^*(k_x)
  = -i a_\mp \psi_\mp(-k_x).
 \label{rel}
\end{equation}

Assume a semi-infinite plane $y > 0$ with the hard wall boundary at $y = 0$.
The eigenstate of the boundary problem will contain both the incident wave
$\psi_-$ and reflected wave $\psi_+$ with some amplitudes. In order to fulfill
the boundary condition $\psi|_{y = 0} = 0$, we have to add a third eigenfunction
of the 2D problem exponentially decaying in the bulk $y > 0$. This solution has
the form
\begin{gather}
 \psi_0
  = \begin{pmatrix}
      1 \\ a_0
    \end{pmatrix} e^{i k_x x - \sqrt{1 - B^2 (E^2 - k_x^2)} y / B},
 \\
 a_0
  = \frac{B k_x - \sqrt{1 - B^2 (E^2 - k_x^2)}}{1 + B E}.
\end{gather}
The scattering state can now be directly constructed:
\begin{gather}
 \psi
  = \psi_- + r\, \psi_+ + r_0\, \psi_0,
 \\
 r
  = \frac{a_- - a_0}{a_0 - a_+},
 \qquad
 r_0
  = \frac{a_+ - a_-}{a_0 - a_+}.
\end{gather}
Far from the boundary, the solution $\psi_0$ decays and the factor $r$ yields
the reflection amplitude.

If the boundary condition preserve time-reversal symmetry, the following state
would be another legitimate solution of the boundary problem at $y \to +\infty$:
\begin{equation}
 \sigma_y \psi^*
  = \sigma_y (\psi_- + r\, \psi_+)^*.
\end{equation}
Applying relations (\ref{rel}), we obtain
\begin{align}
 \sigma_y \psi^*
  &= \sigma_y \psi_-^* + r^* \sigma_y \psi_+^* \notag \\
  &= -i a_+ \psi_+(-k_x) -i a_- r^* \psi_-(-k_x) \notag \\
  &= -i a_- r^* \left[
       \psi_-(-k_x) + \frac{a_+}{a_- r^*}\, \psi_+(-k_x)
     \right].
\end{align}
Now we compare the prefactor in front of $\psi_+(-k_x)$ with the reflection
amplitude at $-k_x$. This will indicate the degree of time-reversal symmetry
breaking during one scattering off the boundary. Actually, the following
identity holds:
\begin{equation}
 \frac{a_+}{a_- r^*(k_x)}
  = - r(-k_x).
\end{equation}
The sign here is just opposite to what one would expect from the time-reversal
invariance. Thus the symmetry is maximally violated.

\section{Relativistic corrections at the bottom of the band}
\label{App:bottom}

At energies close to the bottom of the band, such that $E = M + \epsilon$ with
$\epsilon \ll M$, the BHZ Hamiltonian acquires an approximate orthogonal
time-reversal symmetry, see Sec. \ref{s3.2}. 
It is related to the fact that the spin of the electron
is almost aligned with the $z$ axis and very weakly depends on the momentum.
The Hamiltonian can be approximated as $h \approx \mathbf{k}^2/2M$ (in this
Appendix we assume $A = 1$ and $B$ = 0) up to small relativistic corrections.
The situation is completely analogous to the non-relativistic Shr\"odinger
approximation to the Dirac Hamiltonian. Below we derive these relativistic
corrections for the upper block of Eq.\ (\ref{1}).

Let us write explicitly the two-component Dirac equation with an external
potential $V$:
\begin{align}
 (M + V) u + k_- v
  &= (M + \epsilon) u, \\
 k_+ u - (M - V) v
  &= (M + \epsilon) v. 
\end{align}
Here $k_\pm = k_x \pm i k_y$. We now express $v$ from the second equation,
\begin{equation}
 v
  = (2M + \epsilon - V)^{-1} k_+ u,
 \label{v}
\end{equation}
and use it to obtain a single second order equation for the component $u$,
\begin{equation}
 \big[V + k_- (2M + B k^2 + \epsilon - V)^{-1} k_+ \big] u
  = \epsilon u.
 \label{su}
\end{equation}
In order to recast it in the form of an equivalent one-component Schr\"odinger
equation, we have to correct the normalization of the wave function. The initial
spinor function is normalized according to $|u|^2 + |v|^2 = 1$. We rescale the
component $u$ as
\begin{equation}
 u
  = \left(
      1 - \frac{k^2}{8M^2}
    \right) \phi.
 \label{uphi}
\end{equation}
With the help of Eq.\ (\ref{v}), this yields $|\phi|^2 = 1 + O(k^4)$. We now
substitute $u$ from Eq.\ (\ref{uphi}) into Eq.\ (\ref{su}) and expand in powers
of $k$. In order to cancel the terms $\sim \epsilon k^2$ we have to multiply
both sides of Eq.\ (\ref{su}) by $(1 - k^2/8 M^2)$. This finally yields the
Schr\"odinger equation $h \phi = \epsilon \phi$ with the Hamiltonian
\begin{equation}
 h
  = \frac{k^2}{2M} - \frac{k^4}{8M^3}
    + V + \frac{k_- V k_+}{4 M^2} - \frac{k^2 V + V k^2}{8 M^2}.
 \label{hrel}
\end{equation}
First two terms represent the expansion of the relativistic dispersion relation
$\epsilon = \sqrt{M^2 + k^2} - M$ in powers of $k$. The last three terms appear
due to external potential $V$. In an analogous calculation for the full 4D
Dirac Hamiltonian, the last two terms of the Hamiltonian would correspond to
spin-orbit scattering. In our 2D case, non-relativistic wave function is just
the scalar $\phi$ without any spin structure. The ``spin-orbit'' scattering in
this case can be rewritten as the interaction with a fictitious magnetic field.
Let us introduce the vector potential according to
\begin{equation}
 a_x
  = -\frac{\nabla_y V}{4eM},
 \quad
 a_y
  = \frac{\nabla_x V}{4eM}.
\label{vector}
\end{equation}
This allows us to represent the Hamiltonian as
\begin{equation}
 h
  = \frac{(\mathbf{k} - e \mathbf{a})^2}{2M} - \frac{k^4}{8M^3}
    + V - \frac{e^2 \mathbf{a}^2}{2 M} - \frac{e b}{2 M}.
\label{heff}
\end{equation}
The effective magnetic field 
$$b=\frac{\nabla^2 V}{4 e M}$$ 
leads to breaking of the approximate
``orthogonal'' time-reversal symmetry with the rate $1/\tau_A$ as calculated in
Sec. \ref{s3.2}, see Eqs. (\ref{tauA})-(\ref{Saresult}).


\begin{thebibliography}{12}
\bibitem{footnoteWL} The term 
$$\ln\left( \frac{\tau}{\tau_\phi}+\frac{\tau}{\tau_{SO}}
     +\frac{\tau}{\tau_A +\tau_\Delta}\right)$$
corresponding to the pair of symmetries $T_{xx}$ and $T_{yy}$ 
is written in the form that describes correctly all the asymptotics
and crossovers, except possibly for the ``triple point''
$\text{min}\left[\tau_\phi,\tau_{SO}\right]\sim \tau_A\sim \tau_\Delta$,
when both  $T_{xx}$ and $T_{yy}$ symmetries are broken on the same scale.
Note that this term contains a sum of symmetry-breaking times (not rates).
With the same accuracy one can use an alternative (less compact) expression
\begin{eqnarray*}
\ln\left( \frac{\tau}{\tau_\phi}+\frac{\tau}{\tau_{SO}}
     +\frac{\tau}{\tau_A}\right)+\ln\left( \frac{\tau}{\tau_\phi}+\frac{\tau}{\tau_{SO}}
     +\frac{\tau}{\tau_\Delta}\right) 
\\ 
- \ln\left( \frac{\tau}{\tau_\phi}+\frac{\tau}{\tau_{SO}}
     +\frac{\tau}{\tau_A} +\frac{\tau}{\tau_\Delta}\right)
\end{eqnarray*}
which contains a sum of symmetry-breaking rates.
In order to rigorously describe the above ``triple-point crossover'', 
one should perform a microscopic 
calculation.
\end{thebibliography}
\end{document}